# Design of A Low Power Low Voltage CMOS Opamp

Ratul Kr. Baruah

Department of Electronics and Communication Engineering, Tezpur University, India

ratulkr@tezu.ernet.in

## ABSTRACT

*In this paper a CMOS operational amplifier is presented which operates at 2V power supply and 1uA input bias current at 0.8 um technology using non conventional mode of operation of MOS transistors and whose input is depended on bias current. The unique behaviour of the MOS transistors in subthreshold region not only allows a designer to work at low input bias current but also at low voltage. While operating the device at weak inversion results low power dissipation but dynamic range is degraded. Optimum balance between power dissipation and dynamic range results when the MOS transistors are operated at moderate inversion. Power is again minimised by the application of input dependant bias current using feedback loops in the input transistors of the differential pair with two current substractors.  In comparison with the reported low power low voltage opamps at 0.8 um technology, this opamp has very low standby power consumption with a high driving capability and operates at low voltage. The opamp is fairly small (0.0084 $mm^2$) and slew rate is more than other low power low voltage opamps reported at 0.8 um technology [1,2]. Vittoz at al [3] reported that slew rate can be improved by adaptive biasing technique and power dissipation can be reduced by operating the device in weak inversion. Though lower power dissipation is achieved the area required by the circuit is very large and speed is too small. So, operating the device in moderate inversion is a good solution. Also operating the device in subthreshold region not only allows lower power dissipation but also a lower voltage operation is achieved.*

## 1. INTRODUCTION

Silicon technology continues to scale to ever smaller transistor sizes to reach the market need to include more and more transistors in DRAM and faster processing of microprocessor. While such scaling is of great benefit to digital CMOS circuit behavior, analog devices are sometimes hampered by smaller device sizes and lowered supply voltage. As the transistor lengths decrease in size, the channel modulation has a greater effect and drain current increases more with a larger $V_{DS}$. To develop efficient portable electronic equipment the semiconductor industry has pushed the circuit designers towards low voltage power supply and low power consumption of circuits [4]. Though new smaller size process technologies offer opportunities to operate at higher frequencies consuming less power, for analog circuits, this fact partially applies since it is often the case that additional current is needed to keep the same performance when the power supply voltage is decreased.  Power dissipation in a circuit can be reduced by reducing either supply voltage or total current in the circuit or by reducing the both. As the input current is lowered though power dissipation is reduced, dynamic range is degraded. As the supply voltage decreases, it also becomes increasingly difficult to keep transistors in saturation with the voltage headroom available [5]. Another concern that draws from supply voltage scaling is the threshold of the transistor. A decrease in supply voltage without a similar decrease in threshold voltage leads to biasing issues. Thus, a typical analog design techniques are needed in order to face the above issues.

One important design aspect to challenge the low voltage operation is operating the analog devices in subthreshold region. In this paper a CMOS operational amplifier is presented which operates at 2V power supply and 1uA input bias current at 0.8 um technology using non conventional mode of operation of MOS transistors and whose input is depended on bias current. The unique behaviour of the MOS transistors in subthreshold region not only allows a designer to work at low input bias current but also at low voltage. For CMOS analog circuits, when the transistors operate in weak inversion region, $g_m/I_D$ reaches the maximum [6]; hence the minimum power consumption can be achieved due to the small quiescent current at the expense of large silicon area and slow speed. When MOS transistors operate in strong inversion, however, although good frequency response and small area are obtained, non-optimum larger power is consumed, and $V_{DS}$ (sat) is high. So, it seems a painful tradeoff question between high and low power [7]. The best tradeoff among area, power and speed is achieved when the transistors work in moderate inversion region [8,9,10]. To reduce the current in the circuit the input is made bias dependant using feedback loops in the input transistors of the differential pair with two current substractors.

The reported low power low voltage amplifiers using classical schemes [11,12,13] have good small signal characteristics but their slew rate is small. By using the above techniques slew rate is improved, as well as lower power dissipation is achieved. Also as the transistors are operated at weak and moderate inversion of MOS transistor, the opamp operates at low power as well as low voltage.

## 2. PRINCIPLE

The basic transconductance amplifier used is shown in fig. 1. When a voltage is applied across the inputs of the opamp the currents $I_1$ and $I_2$ become different. The bias current of the amplifier is made signal dependent by adding an additional current source to the main tail current source which realizes $I_{BIAS} + A.|I_1 - I_2|$, A is current feedback factor (normally A<1).

The additional current source is realized by two substractors (Fig. 2). When $I_2>I_1$, the substractor draw a current $A.|I_2 - I_1|$, otherwise it keeps drawing zero current.

Fig. 3 shows the implementation of adaptive biasing into the OTA of fig 1. If there is no disturbance at the virtual ground, $I_1=I_2$ and the total bias current is thus $I_{BIAS}$. When a signal is applied, the total bias current become $I_{BIAS} + A.|I_1 - I_2|$.

Maximum total current available through M1 occurs when M2 is off. Positive feedback exits through the loop M1, M3, M11, M18. Initially, when M2 shuts off, the current in M1, M3 is $I_{BIAS}$. This is mirrored in M11, M16 and the current through M18 is thus $A.I_{BIAS}$. At a particular instant of time, the tail current, which flows through M1, is now $I_{BIAS} + A.I_{BIAS}$, Provided that MOSFETS M1,M3,M18 remains in saturation, this current circles back around the positive feedback loop and increases by A. This continues, resulting in a final or total tail current of

$$I_{TOTAL} = I_{BIAS}.(1+A+A^2+A^3+\ldots) \quad (1)$$

If A<1, eq$^n$ (1) can be written as

$$I_{TOTAL} = \frac{I_{BIAS}}{1-A} \quad (2)$$

If A=0, ie M18 does not exist, then there is no adaptive biasing and the total tail current is $I_{BIAS}$.

Since the tail current limits the slew rate, when the amplifier is driving a capacitive load, making A equal to one eliminates slew rate limitation while at the same time not increasing static power dissipation.

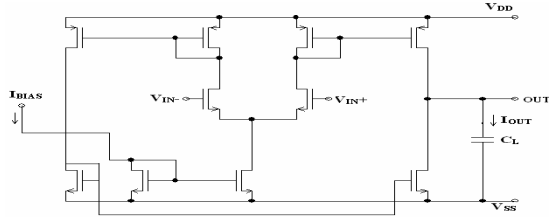

Fig. 1 The basic transconductance amplifier

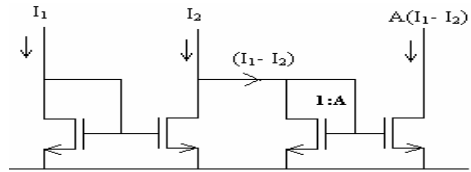

Fig 2 Current Substractor realizing $A.(I_2-I_1)$

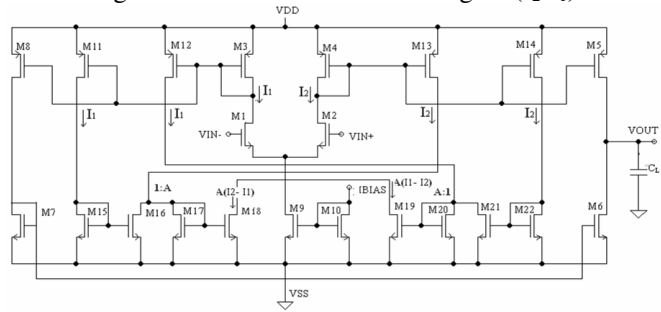

Fig 3 Block diagram of low voltage low power opamp

Available Output current

By using the current formula for the weak inversion region [14,15] the following equations can be derived.

$$I_1 = I_2 . \exp(V_{in}/(n . V_T)) \quad (3)$$

$$I_1 = \frac{I_{BIAS} . \exp(V_{in}/(n . V_T))}{(A+1)-(A-1).\exp(V_{in}/(n . V_T))} \quad (4)$$

The current flowing in the load capacitor, $I_{OUT}$ is b times the difference between $I_1$ and $I_2$.

$$I_{OUT} = \frac{I_{BIAS} \cdot (\exp(V_{in}/(n \cdot V_T)) - 1)}{(A+1) - (A-1) \cdot \exp(V_{in}/(n \cdot V_T))} \cdot b \qquad (5)$$

b is the ratio between the current in the second and the first stage. $V_{in}$ is the voltage across the inputs of the amplifier. n is the slope factor in weak inversion. $V_T = kT/q$, thermal voltage.

When the transistor is operated in moderate inversion the equation is somewhat complicated for hand calculation. BSIM3v3 model provides one equation for all regions with good accuracy [7,8].

### 2.1. Simulated Results

Circuit operating conditions

Operating Temperature = room temperature
Power supply= 2V, Input bias current =1 uA, $C_{load}$ =5 pf

The simulating specs of the opamp (after post parasitic simulation) are listed in table 1.

TABLE I

| Parameter | Simulated Value |
| --- | --- |
| DC gain | 61.11dB |
| UGB | 1.09 $MH_Z$ |
| Input Bias Current | 1uA |
| Total Power Dissipation | 16.8 uW |
| Phase Margin | 61° |
| ICMR | 0.16 V-1.87 V |
| Input Offset Voltage | 0.042144 mV |
| Output Swing | 0.115 V-1.88 V |
| Common Mode Gain | 0.0084673dB |
| CMRR | 60.3 dB |
| Slew Rate(Rise/Fall) | 2.84 V/us, 2.85 V/us |
| PSRR(+/-) | 108dB, 110dB |
| Settling Time(Rise/Fall)(1%) | 0.9 us / 0.92 us |
| Noise | 1.2 uV/$\sqrt{HZ}$ (At 1 $MH_Z$) |
| Area | 0.0084 $mm^2$ |

### 2.2. Simulated Curves

Fig. 4 shows the DC transfer curve, DC output voltage versus DC input voltage in open loop configuration. The offset voltage is very small 0.04 mV, which is desirable.

Fig. 5 shows the Gain curve, DC gain in dB versus frequency in Hz (in log scale) & Phase Margin of opamp in open loop. The DC gain of the opamp is 61.11dB, Phase margin, $61^O$ and Unity gain bandwidth is 1.09 $MH_Z$ which is good for low power low voltage opamp. Fig. 6 shows the output voltage swing of the opamp over which the output voltage can swing while still maintaining same high gain characteristics. The output swing of the opamp (0.115 V-1.88 V) is excellent. It is very near to the supply rails. Fig 7 shows the Input common mode range (ICMR) of the opamp over which the input common mode signal can vary. The ICMR of the opamp is 0.16 V-1.87 V, which is excellent and very near the supply rails. Fig 8 shows the Slew Rate of the opamp which gives the output voltage rate. Settling time, the time needed for the output of the opamp to reach a final value when excited by a small signal can be found from the same characteristic. The Slew Rate of the opamp is 2.85 V/uS, which is quite good as compared to other low power oamps. The Settling Time of the opamp is 0.9 us. Fig 10 shows $I_{OUT}/I_{BIAS}$ Vs $V_{IN}/nV_T$ & supply current vs $V_{IN}$ of opamp for A=1. If A>1 , the maximum load current become unlimited, so that the amplifier will never slew. The maximum possible output current will be determined by the beta's of the transistors and the supply voltage.

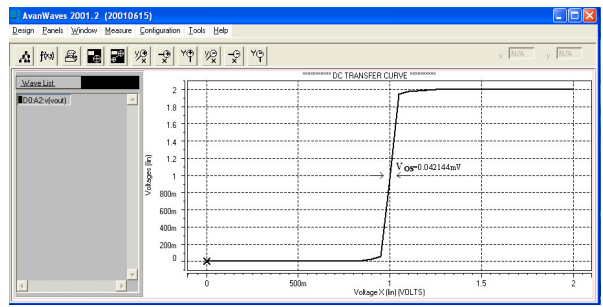

Fig. 4 Open loop Transfer characteristics

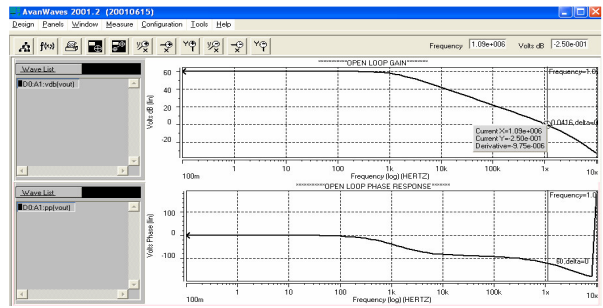

Fig. 5 Gain & Phase Margin of opamp

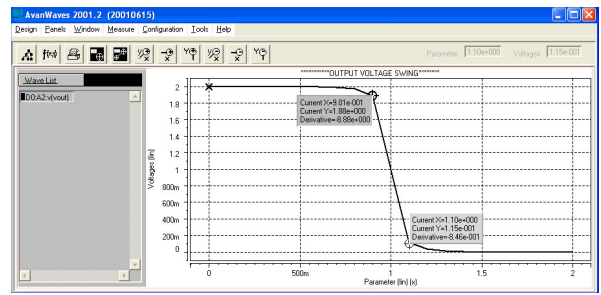

Fig: 6 Output Voltage Swing of opamp

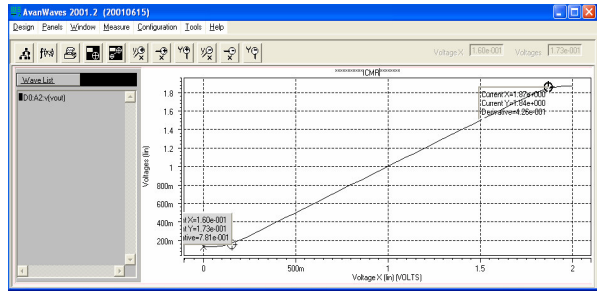

Fig 7 ICMR of the opamp

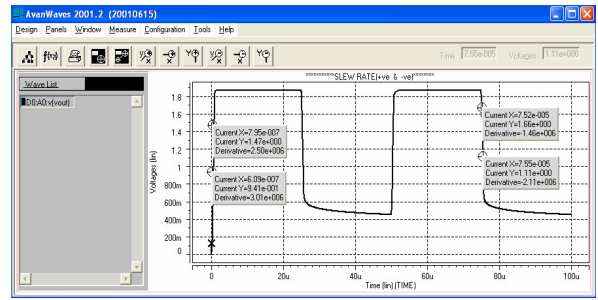

Fig 8  Slew Rate (+ve & -ve) of opamp

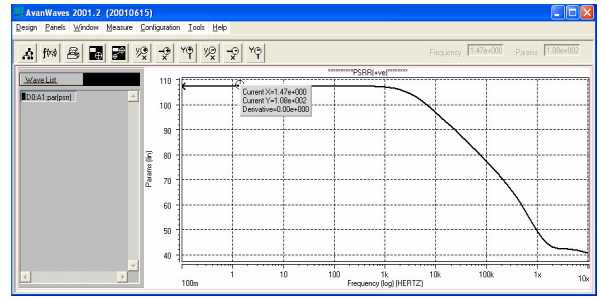

Fig 9  PSRR of opamp

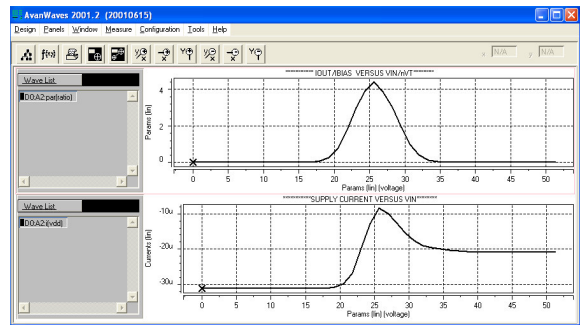

Fig 10  $I_{OUT}/I_{BIAS}$ Vs $V_{IN}/nV_T$ & supply current vs $V_{IN}$ of  opamp

## 2.7. Layout

The layout of my circuit shown in Fig [11] is done using professional software Mentor Graphics. The layout has area 0.0084 mm$^2$, which is fairly small as compared to other low power opamps [6]. The layout is both DRC and LVS clean. Parasitic extraction and Post simulation is performed successfully.

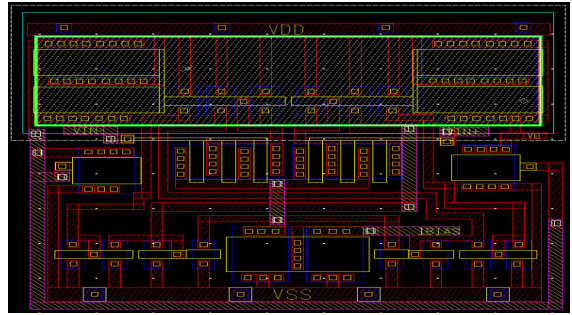

Fig. 11  Layout of 2V 1uA opamp

## 3. CONCLUSIONS

The amplifier presented in this paper operates at weak and moderate inversion and regulates its bias current. When a signal is applied the current in the amplifier increases so that these amplifiers have very high driving current. As the opamp works at weak and moderate inversion, low power as well as low voltage can be achieved. Moreover, best tradeoff among area, power and speed is achieved when the transistors work in moderate inversion region. Its slew rate is higher than reported low power low voltage amplifiers at 0.8 um technology. Also its size is fairly small. The simulations and layout is done using professional software Mentor Graphics.

## REFERENCES

[1]     H. A. Aslanzadeh, S. Mehrmanesh, M.B. Vahidfar, A.Q. Safarian R.Lotfi, "A 1-V 1-mW High-Speed Class AB  Operational Amplifier for High-Speed Low Power Pipelined A/D Converters using "Slew Boost" Technique" , ISLPED'0*3*, August 25–27, 2003, Seoul, Korea.

[2]     Troy S.,H Yoshizawa,"A 0.9V,0.5uARail o Rail CMOS operational Amplifier", IEEE 2001 customIntegrated circuits  conference, 2001.

[3]     M.G.Degrauwe, J.Rijmenants,E.A Vittoz, "Adaptive biasing CMOS  amplifiers", IEEE journal of solid state  Electronics,Vol sc 17,No 3,June 1982.

[4]     The International Technology Roadmap for semiconductors, ITRS.

[5]     R.Gonzalez, B.M. Gordon, M.A Horowitz, "Supply and Threshold voltage scaling for low power CMOS", IEEE journal of solid state  Electronics,Vol sc 32,No 8,June 1997.

[6]     R.L O.Pinto, M.C. Schneider,C.G. Montoro, "Optimum Design of MOS Amplifiers", XII Brazilian Symposium on  Integrated Circuits and Systems Design, September 29-October 01,1999

[7]     C.H.Lin and M. Ismail, "A Low Voltage Low power CMOS opamp with Rail-to-Rail Input/Output",  , IEEE journal of solid state  Electronics,Vol sc 32,No 6,June 1997.

[8]     Amin Shameli, Payam Heydari, "A Novel Power Optimization Technique for Ultra-Low Power RFICs", ISLPED'06, October 4–6, 2006, Tegernsee, Germany.

[9]     Fabrice Guigues, Edith Kussener, Benjamin Duval, and Herv´e Barthelemy "Moderate Inversion: Highlights for Low Voltage Design", PATMOS 2007, LNCS 4644, pp. 413–422, 2007.

[10]    Binkley, D., Hoper, C., Trucker, S., Moss, B., Rochelle, J., Foty, D., "A CAD methodology for optimizing transistor current and sizing in analog CMOS design", IEEE Transactions on Computer-Aided Design of Integrated Circuits and Systems 22(2), 225–237 (2003).

[11]    F.-Krummenacher,JL Zufferey,"High gain cmos cascade opamp",Electronics Letter, Vol 7, Feb 1981.

[12]    F.-Krummenacher,"High gain CMOS OTA for micropower SC filters",Springer-Verlag, 1985, ch. 4.

[13]    W.Steinhagen andW L Engl," Design of integrated analog CMOS Circuits-A multichannel telemetry transmitter ",IEEE JSSE,VOLSC 13, Dec '78.

[14]    E.Vittoz,J.Felirath,"CMOS ANALOG Integrated circuits based on weak inversion operation", IEEE journal of solid state Electronics,Vol sc 12No 3,June 1977.

[15]    Enz, C.C., Krummenacher, F., Vittoz, E.A.,"An analytical MOS transistor model valid in all regions of operation and dedicated to low-voltage and low-current", Analog Integrated Circuits and Signal Processing 8, 83–114 (1995)